\documentclass{article}

\usepackage{arxiv}

\usepackage[utf8]{inputenc} 
\usepackage[T1]{fontenc}    
\usepackage{hyperref}       
\usepackage{url}            
\usepackage{booktabs}       
\usepackage{amsfonts}       
\usepackage{nicefrac}       
\usepackage{microtype}      
\usepackage{lipsum}
\usepackage{bm}
\usepackage{amsmath, amssymb}
\usepackage{comment}
\usepackage{multirow}
\usepackage{graphicx} 

\usepackage[whole]{bxcjkjatype}

\title{Schr\"{o}dinger Risk Diversification Portfolio}

\author{
  Yusuke Uchiyama\\
  MAZIN, Inc.\\
Amazing Building 3F, 3-29-14 Nishiasakusa, Taito-ku, Tokyo 111-0035, Japan\\  \texttt{uchiyama@mazin.tech} \\
   \And
  Kei Nakagawa\\
  NOMURA Asset Management Co. Ltd,\\
  2-2-1 Toyosu, Koto-ku, Tokyo 135-0061\\
  \texttt{kei.nak.0315@gmail.com} \\
}

\begin{document}
\maketitle

\begin{abstract}
The mean-variance portfolio that considers the trade-off between expected return and risk has been widely used in the problem of asset allocation for multi-asset portfolios. 
However, since it is difficult to estimate the expected return and the out-of-sample performance of mean-variance portfolio is poor, risk-based portfolio construction methods focusing only on risk have been proposed, and are attracting attentions mainly in practice.
In terms of risk, asset fluctuations that make up the portfolio are thought to have common factors behind them, and principal component analysis, which is a dimension reduction method, is applied to extract the factors.
In this study, we propose the Schr\"{o}dinger risk diversification portfolio as a factor risk diversifying portfolio using Schr\"{o}dinger principal component analysis that applies the Schr\"{o}dinger equation in quantum mechanics. 
The Schr\"{o}dinger principal component analysis can accurately estimate the factors even if the sample points are unequally spaced or in a small number, thus we can make efficient risk diversification.
The proposed method was verified to outperform the conventional risk parity and other risk diversification portfolio constructions.

\end{abstract}


\section{Introduction}The mean-variance portfolio, which takes into account the trade-off between expected return and risk, has been used in the problem of asset allocation for portfolios consisting of multiple assets.  
Since its proposal by Markowitz~\cite{markowitz1952portfolio}, the mean-variance portfolio has long been an important framework in investment decision making.  
The reason is that the mean-variance portfolio is easy to handle in practice because the optimization problem can be expressed as a quadratic programming problem. 
The mean-variance portfolio requires the expected return, the variance, and the correlation among the assets and uses them as inputs to determine the asset allocation by considering the trade-off between the expected return and the variance of the portfolio under various investment constraints.  

However, the mean-variance portfolio has several drawbacks.  
The most frequently pointed out problem is that it is quite difficult to predict the exact value of the expected returns as an input in the first place, thus the out-of-sample performance is too poor~\cite{merton1980estimating}. 
Market uncertainty is increasing year by year, as exemplified by the Lehman shock and the recent Covid-19 shock, and the estimating expected returns is becoming more and more difficult under such market conditions.  
Another problem is that the mean-variance portfolio obtained as a result of optimization differs greatly for small changes in expected returns~\cite{michaud1989markowitz}.

Therefore, risk-based portfolio construction methods that focus only on risk without expected returns have been proposed and are attracting attention mainly in practical applications.  
The minimum risk portfolio~\cite{clarke2011minimum,rockafellar2000optimization}, the risk parity portfolio~\cite{qian2005risk,maillard2010properties}, and the maximum diversification portfolio~\cite{choueifaty2008toward} are typical examples, and risk-based portfolios that are further developed from them are proposed~\cite{uchiyama2019complex,nakagawa2020go,nakagawa2020RM,uchiyama2020tplvm,nakagawa2020RM}.
In fact, in various empirical studies and backtests on equity portfolios and asset allocations, these risk-based portfolio construction methods perform better than traditional mean-variance portfolios and capitalization-weighted portfolios~\cite{poddig2012robustness}.
Also, unlike the mean-variance portfolio, the performance of the risk-based portfolio is less affected by the estimation accuracy of the risk~(covariance matrix)~\cite{nakagawa2018risk}.

\par
On the other hand, asset fluctuations that make up a portfolio are thought to have common factors behind them.
Principal component analysis, which is a dimension reduction method, is applied to extract these factors.
Based on this idea, the factor-based risk parity portfolio, which allocates risk equally to factors extracted by principal component analysis, has been proposed as a risk-based portfolio construction method~\cite{meucci2009managing}.  
In contrast to the usual risk parity portfolio, which allocates risk equally on assets, the factor-based risk parity portfolio is called a risk diversification portfolio.  

In addition, for the purpose of extracting dynamic risk factors, a complex-valued risk diversification portfolio is proposed, which allocates risk equally to the factors extracted by Hilbert principal component analysis~\cite{uchiyama2019complex}, and a quaternion risk diversification portfolio is further developed~\cite{sugitomo2020quaternion}.

\par
In this study, we propose the Schr\"{o}dinger Risk Diversification Portfolio as a novel risk diversification portfolio using Schr\"{o}dinger principal component analysis~\cite{PhysRevE.104.025307}, which is an application of the Schr\"{o}dinger equation in quantum mechanics. The Schr\"{o}dinger principal component analysis can estimate the principal components and cross-correlations with high accuracy even when the data are unequally sampled or a small number.
Therefore, we expect to estimate factors accurately and enable efficient risk diversification.
We compare the proposed method with existing risk diversification portfolios to verify its effectiveness using real asset data.

\section{Preliminary}
\subsection{Gaussian Filed and Schr\"{o}dinger Equation}
%
Suppose, given a scalar field ${\phi}:\mathbb{R}^n\to{\mathbb{R}}$ for spatial coordinates $\bm{x}\in{\mathbb{R}^d}$.

When the quantity of the field ${\phi}(\bm{x}_1), {\phi}(\bm{x}_2), {\cdots}, {\phi}(\bm{x}_N)$ corresponding to any $N$ coordinate vectors $\bm{x}_1, \bm{x}_2, {\cdots}, \bm{x}_N$ is a random variable following an $N$-dimensional Gaussian distribution, ${\phi}({\cdot})$ is called a Gaussian process.


In particular, ${\phi}({\cdot})$ is also called a Gaussian field because it is a field quantity with respect to the spatial coordinate $\bm{x}$.  
By subtracting the mean value $\mathbb{E}[{\phi}(\bm{x})]$ from the Gaussian field, we can assume to set $\mathbb{E}[{\phi}(\bm{x})]=0$ without loss of generality.

Then, the positive definite symmetric function given as follows is called the covariance function or kernel function.  
\begin{equation}
k(\bm{x},\bm{x}') = \mathbb{E}[{\phi}(\bm{x}){\phi}(\bm{x}')]
\end{equation}
%


Thus, for any $N$ coordinate vector $\bm{x}_1, \bm{x}_2, {\cdots}, \bm{x}_N$, the kernel function gives the covariance matrix of the $N$-dimensional Gaussian distribution.

\begin{equation}
K = 
\left[
    \begin{array}{cccc}
      k(\bm{x}_1,\bm{x}_1) & k(\bm{x}_1,\bm{x}_2) & \ldots & k(\bm{x}_1,\bm{x}_N) \\
      k(\bm{x}_2,\bm{x}_1) & k(\bm{x}_2,\bm{x}_2) & \ldots & k(\bm{x}_2,\bm{x}_N) \\
      \vdots & \vdots & \ddots & \vdots \\
      k(\bm{x}_N,\bm{x}_1) & k(\bm{x}_N,\bm{x}_2) & \ldots & k(\bm{x}_N,\bm{x}_N)
    \end{array}
  \right]
\end{equation}
%


For the covariance function of a Gaussian field, there exists a real function $A({\cdot})$ and a positive definite symmetric matrix ${\Sigma}$ that satisfies 
\begin{align}
k(\bm{x},\bm{x}') &= \sqrt{A(\bm{x})A(\bm{x}')} \nonumber \\
&{\times} \exp{\left[-\frac{1}{2}(\bm{x}-\bm{x}')^{T}{\Sigma}^{-1}(\bm{x}-\bm{x}')\right]}.
\end{align}

%
On the other hand, for an integral operator with a kernel function as its integral core, there exist the following eigenvalue problem
\begin{equation}
\int_{{\Omega}_{\bm{x}}}k(\bm{x}, \bm{x}'){\varphi}_i(\bm{x}')d\bm{x}' = {\lambda}_i{\varphi}_i(\bm{x}'),
\label{eq:eigen_eq}
\end{equation}
%
where ${\lambda}_i$ and ${\varphi}_i$ are $i$-th eigenvalue and eigenfunction, respectively, which satisfy the orthogonality condition
\begin{equation}
\int_{{\Omega}_{\bm{x}}}{\varphi}_i(\bm{x}){\varphi}_j(\bm{x})d\bm{x} = {\delta}_{i,j}
\end{equation}
%
where ${\delta}_{i,j}$ is the Kronecker delta.  
Due to the positive symmetry of the kernel function, the eigenvalues are positive real numbers.  
The subscripts $i$ and $j$ of the eigenvalues and eigenfunctions are taken to satisfy ${\lambda}_1\geq{{\lambda}_2}\geq{{\cdots}}\geq{0}$.  
Since the Eq. (\ref{eq:eigen_eq}) corresponds to the infinite-dimensional vector of the eigenvalue problem of the covariance matrix, the eigenfunction expansion censored by an appropriate finite number of eigenfunctions corresponds to the principal component analysis.  \par
%
The eigenvalue problem of the Schr\"{o}dinger equation in quantum mechanics is given as follows:
\begin{equation}
-\frac{1}{2}{\Delta}_{{\Sigma}_{\bm{m}}}{\psi}(\bm{x})+V(\bm{x}){\psi}(\bm{x}) = E{\psi}(\bm{x}).
\label{eq:SEQ}
\end{equation}
%
Here, ${\psi}({\cdot})$ is the wave function and ${\Delta}_{{\Sigma}_{\bm{m}}}$ is the second-order partial differential operator weighted by the positive definite symmetric matrix ${\Sigma}_{\bm{m}}$.
\begin{equation}
{\Delta}_{{\Sigma}_{\bm{m}}}=\sum_{i,j}\frac{{\partial}}{{\partial}x_i}({\Sigma}_{\bm{m}})_{i,j}\frac{{\partial}}{{\partial}x_j}
\end{equation}
%
The real function $V({\cdot})$ is a potential function and $E$ is an energy eigenvalue.  
The linear operator appearing on the left-hand side of Eq. (\ref{eq:SEQ}) is also called the Hamilton operator or Hamiltonian.
\begin{equation}
H = -\frac{1}{2}{\Delta}_{{\Sigma}_{\bm{m}}}+V(\bm{x})
\end{equation}
%
Since the Hamiltonian operator is a Hermite operator, the energy eigenvalues are non-negative real values.  
\subsection{Schr\"{o}dinger PCA}
%
The integral equation in Eq. (\ref{eq:eigen_eq}) and the Schr\"{o}dinger equation expressed in Eq. (\ref{eq:SEQ}) agree to the second order of accuracy according to the following correspondence:
\begin{align}
{\phi}(\bm{x}) \; &{\Leftrightarrow} \; {\psi}(\bm{x}) \\
-A(\bm{x}) \; &{\Leftrightarrow} \; V(\bm{x}) \\
A(\bm{x}){\Sigma} \; &{\Leftrightarrow} \; {\Sigma}_{\bm{m}}.
\end{align}
%
Therefore, we can perform principal component analysis by expanding the original Gaussian field by the eigenfunctions of the Schr\"{o}dinger equation and then terminating the expansion at an appropriate point.  This is called Schr\"{o}dinger principal component analysis~(PCA)\cite{PhysRevE.104.025307}.  
Since ordinary principal component analysis can be performed using a finite data set, it is known that the estimation accuracy deteriorates when only a few samples are available.  
On the other hand, since Schr\"{o}dinger PCA does not depend on the number of data samples, we can obtain stable estimation results within the range of approximation accuracy up to the second order.

\section{Method:Schr\"{o}dinger Risk Diversification Portfolio}
%

Schr\"{o}dinger PCA considers the quantity of a field, that is, the quantity that fluctuates at each point of spatial coordinates.  
On the other hand, the asset allocation problem considers a system in which the prices of individual assets fluctuate, which cannot be mapped to spatial coordinates.  
Therefore, we apply Schr\"{o}dinger PCA to multivariate time series consisting of multiple assets price fluctuations by conveniently mapping the $i$th asset to the spatial coordinate $x_i$.  
We formulate the risk diversification portfolio constructed using the results of Schr\"{o}dinger PCA as the Schr\"{o}dinger risk diversification portfolio.

\par
%
The procedure for constructing a Schr\"{o}dinger risk diversification portfolio is as follows:

\begin{enumerate}
    \item 
    For the series of $T$-period returns of $N$ assets, $\{r_{t,i}\}(i=1,2,{\cdots},N;t=1,2,{\cdots},T)$, calculate 
    \begin{align}
    &{\mu}_{i} = \frac{1}{T}\sum_{t=1}^Tr_{t,i}, \\
    &{\sigma}_{i}^2 = \frac{1}{T-1}\sum_{t=1}^T(r_{t,i} - {\mu}_{r,i})^2.
    \end{align}
    %
    %
    \item 
    
    Assume the functional form of the potential function, we perform the substitution $i{\mapsto}j(i)$ such that $V(x_{j(i)})={\sigma}_{i}^2$ and estimate $x_0$ and ${\Delta}x$ in $x_j=x_0+j{\Delta}x$ and the parameters of the potential function.

    \item 
    %
    %
    
    By solving the eigenvalue problem of the Schr\"{o}dinger equation in one spatial dimension, 
    
    \begin{equation}
    -\frac{1}{2}\frac{d^2}{dx^2}{\psi}(x)+V(x){\psi}(x) = E{\psi}(x)
    \end{equation}

  we obtain the pair of eigenvalues and eigenfunctions $\{E_l,{\psi}_l\}(l=1,2,{\cdots},L)$.   Here, $L$ corresponds to the censored order of Schr\"{o}dinger PCA.

    \item 
    %
    %
    %
    %
    Generate a following matrix from the eigenfunctions.
    \begin{equation}
    {\Psi}_{l,i}={\psi}_l(x_i)\;(i=1,2,{\cdots},N;l=1,2,{\cdots},L)
    \end{equation}
    
    For the portfolio weights $w_i\;(i=1,2,{\cdots},N)$, we calculate
    \begin{equation}
    \tilde{w}_l = \sum_{i=1}^N{\Psi}_{l,i}w_i
    \end{equation}

    \item   For each $l$, we obtain $v_l$ from the eigenvalues $E_l$ and $\tilde{w}_l$.
    \begin{equation}
    v_l = \frac{E_l\tilde{w}_l^2}{\sum_{l=1}^LE_l\tilde{w}_l^2}
    \end{equation}
    
    %
    %
    %
    \item We find the optimal weights by maximizing the entropy $\mathcal{H}=-\sum_{l=1}^Lv_l\log{v_l}$ subject to constraints i.e., $\sum_{i=1}^Nw_i=1$.

    %
    %
    %
    %
    %
\end{enumerate}
%
In the case of one-dimensional space, it is known that the Schr\"{o}dinger equation under appropriate potential functions has an analytic solution.

For example, in the case of a harmonic oscillator whose potential function is given by
\begin{equation}
    V(x) = -\frac{1}{2}kx^2.
\end{equation}

Then, $y=x/\sqrt{k}$, ${\lambda}=2E/\sqrt{k}$, we have a second-order ordinary differential equation
\begin{equation}
\frac{d^2{\psi}}{dy^2}+({\lambda}-y^2){\psi}=0
\end{equation}

Solving this equation, we obtain 
\begin{align}
&{\lambda}_l=2l+1, \\
&{\psi}_l(y)=H_l(y)e^{-y^2/2}
\end{align}

Here, the function $H_l({\cdot})$ is a Hermitian polynomial.  
This gives us the eigenvalues and eigenfunctions we wish to obtain.  
There are other potential functions for which analytical solutions can be obtained, and we can use them according to our problem.

%
%
%
\section{Experiment}
In this section, we test the performance of the purposed portfolio construction by comparing with the conventional risk diversifying portfolio constructions, the PCA and HPCA portfolio constructions. 
Data description and the results of performance test are given in the following subsections.

\subsection{Dataset}
To evaluate the performance of each portfolio, we selected six bond futures and six equity futures, as shown in Table \ref{tabl:statistics}. 

All of the daily historical data were corrected during May 2000 to April 2017.
We obtained all of these data from Bloomberg terminals.
The descriptive statistics of the returns of the assets are shown in Table \ref{tabl:statistics}.

\begin{table}[]
\caption{List of Assets.}
\centering
\begin{tabular}{|c|c|c|}
\hline
Ticker & Type          & Description                      \\ \hline
TX1    & Bond Future   & 10 Year T-Note Futures           \\ \hline
XM1    & Bond Future   & Australian 10 Year Bond          \\ \hline
CN1    & Bond Future   & Canadian Government 10 Year Note \\ \hline
RX1    & Bond Future   & Eurex Euro Bund                  \\ \hline
G1     & Bond Future   & Gilt UK                          \\ \hline
JB1    & Bond Future   & Japan 10 Year Bond Futures       \\ \hline
SP1    & Equity Future & S\&P500                          \\ \hline
XP1    & Equity Future & S\&P/ASX 200(Austraria)          \\ \hline
PT1    & Equity Future & S\&P/TSX 60 Index(Canada)        \\ \hline
GX1    & Equity Future & DAX(German)                      \\ \hline
Z1     & Equity Future & FTSE100(UK)                      \\ \hline
NK1    & Equity Future & Nikkei225(Japan)                 \\ \hline
\end{tabular}
\end{table}

\begin{table}[]
\caption{Descriptive statistics of the dataset.}
\centering
\begin{tabular}{|c|r|r|r|r|}
\hline
    & \multicolumn{1}{c|}{Mean {[}\%{]}} & \multicolumn{1}{c|}{Standard Deviation {[}\%{]}} & \multicolumn{1}{c|}{Skewness} & \multicolumn{1}{c|}{Kurtosis} \\ \hline
TY1 & 0.006763                           & 0.395198                                         & -0.28289                      & 4.197417                      \\ \hline
XM1 & 0.00094                            & 0.065218                                         & -0.13666                      & 1.812167                      \\ \hline
CN1 & 0.008609                           & 0.363551                                         & -0.26973                      & 2.366885                      \\ \hline
RX1 & 0.010441                           & 0.352686                                         & -0.54011                      & 3.825943                      \\ \hline
G 1 & 0.003761                           & 0.442155                                         & -5.72552                      & 150.2678                      \\ \hline
JB1 & 0.003133                           & 0.203517                                         & -0.52795                      & 5.763622                      \\ \hline
SP1 & 0.018462                           & 1.207641                                         & 0.202656                      & 12.4827                       \\ \hline
XP1 & 0.019369                           & 1.012393                                         & -0.22977                      & 4.928742                      \\ \hline
PT1 & 0.017505                           & 1.172459                                         & -0.4545                       & 9.33117                       \\ \hline
GX1 & 0.022293                           & 1.489021                                         & 0.245713                      & 6.613814                      \\ \hline
Z 1 & 0.009463                           & 1.184334                                         & -0.01916                      & 6.814881                      \\ \hline
NK1 & 0.012986                           & 1.555291                                         & 0.023165                      & 12.68528                      \\ \hline
\end{tabular}
\end{table}

\subsection{Experimental Settings}

We compare the proposed method Schrodinger Risk diversification Portfolio (SPCA) with the Equal Weight Portfolio (EW), Principal Component risk diversification Portfolio (PCA), and Complex Principal Component Risk Parity Portfolio (HPCA). Each portfolio re-estimates its parameters and rebalances on the 20 days (monthly).

\begin{description}
    \item[EW:] "EW" stands for equally-weighted portfolio~\cite{demiguel2009optimal}. 
    \item[PCA:] "PCA" stands for principal component risk diversification portfolio that invests evenly in principal component risk by performing principal component analysis on return data for the past 250 days~\cite{meucci2009managing}.
    \item[HPCA:] "HPCA" stands for Hilbert transformed principal component risk diversification portfolio that invests evenly in principal component risk by performing principal component analysis on Hilbert transformed return data for the past 250 days~\cite{uchiyama2019complex}. 
    \item[SPCA:] "SPCA" stands for our proposed portfolio by performing Schrodinger principal component analysis on return data for the past 250 days. 
\end{description}
\subsection{Performance Measures}
In evaluating the portfolio strategy, we used the following measures that are widely used in the field of finance~\cite{nakagawa2020RM}.

The portfolio return at time $t$ is defined as
\begin{equation}
    R_t = \sum_{i=1}^n r_{it}w_{it-1} 
\end{equation}
where $r_{it}$ is the return of $i$-th asset at time $t$, $w_{it-1}$ is the weight of $i$-th asset in the portfolio at time $t-1$, and $n$ is the number of assets. 

We evaluated the portfolio strategy using its annualized return (AR), the risk as the standard deviation of return (RISK), and the risk/return (R/R) (return divided by risk) as the portfolio strategy. R/R is the most important performance measure.

\begin{align}
    {\bf AR} &= \frac{250}{T}\sum_{t=1}^T R_t \label{eq_ar}\\
    {\bf RISK} &= \sqrt{\frac{250}{T-1}\times \sum_{t=1}^T(R_t-\mu)^2}\\
    {\bf R/R} &= {\bf AR}/{\bf RISK}
\end{align}

Here, $\mu = (1/T) \sum_{t=1}^T R_t$ is the average return of the portfolio.

We also evaluated the maximum drawdown (MaxDD), which is another widely used risk measure \cite{magdon2004maximum}, for the portfolio strategy.
In particular, MaxDD is defined as the largest drop from an extremum:

\begin{align}
    {\bf MaxDD} &= \min_{k \in [1,T]}\left(0,\frac{W_k^{\mathrm{Port} }}{\max_{j \in [1,k]} W_j^{\mathrm{Port}}}-1\right) \\
    W_k^{\mathrm{Port}} &= \sum_{i=1}^k (1+R_t).
\end{align}

\subsection{Results}
Table~\ref{Perf} summarizes the performance measures of each portfolio.
We can see that SPCA has the highest AR and R/R while HPCA has the lowest Risk. 
Therefore, the proposed method can efficiently earn returns and can also control risks by appropriately diversifying the factor risks.

\begin{table}[]
\centering
\caption{Annual return (AR), risk (RISK) and, risk/return (R/R), maximum drawdown(MaxDD) of the EW, PCA, HPCA and SPCA portfolio.}
\begin{tabular}{|c|r|r|r|r|}
\hline 
      & \multicolumn{1}{c|}{EW} & \multicolumn{1}{c|}{PCA} & \multicolumn{1}{c|}{HPCA} & \multicolumn{1}{c|}{SPCA} \\ \hline 
AR~[\%]    & 3.16                    & 1.58                     & 3.20                      & \textbf{3.84}             \\ \hline
RISK~[\%]  & 6.67                    & 6.31                     & \textbf{5.36}             & 5.59                      \\ \hline
R/R    & 0.47                    & 0.25                     & 0.60                      & \textbf{0.69}             \\ \hline
MaxDD~[\%] & 32.35                   & 26.76                    & \textbf{21.53}            & 34.66                     \\ \hline
\end{tabular}
\label{Perf}
\end{table}

\begin{figure}
  \centering
  \includegraphics[width=0.9\textwidth]{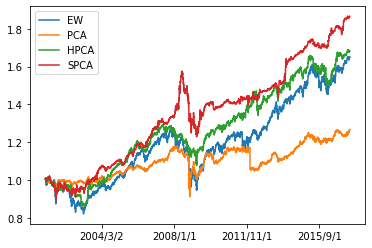}
  \caption{The cumulative returns of the EW~(blue), PCA~(orange), HPCA~(green), and SPCA~(red) portfolio.}
  \label{W}
\end{figure}

In fact, as is shown in Figure~\ref{W}, the cumulative return of the SPCA is stable.

\begin{figure}
  \centering
  \includegraphics[width=0.9\textwidth]{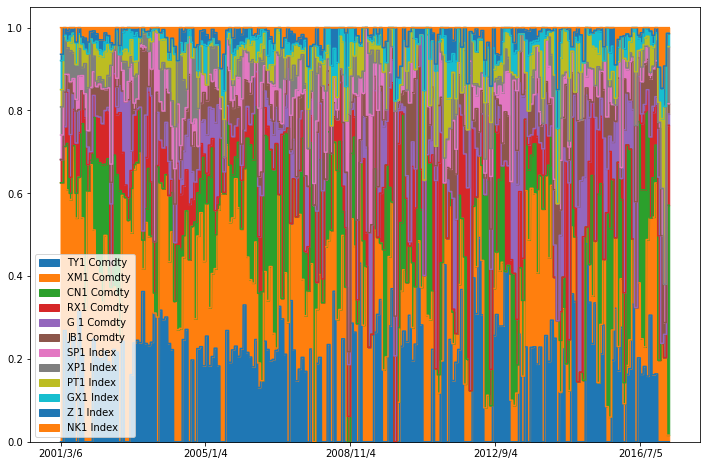}
  \caption{The allocation of the assets in the PCA portfolio.}
  \label{Wgt_PCA}
\end{figure}

\begin{figure}
  \centering
  \includegraphics[width=0.9\textwidth]{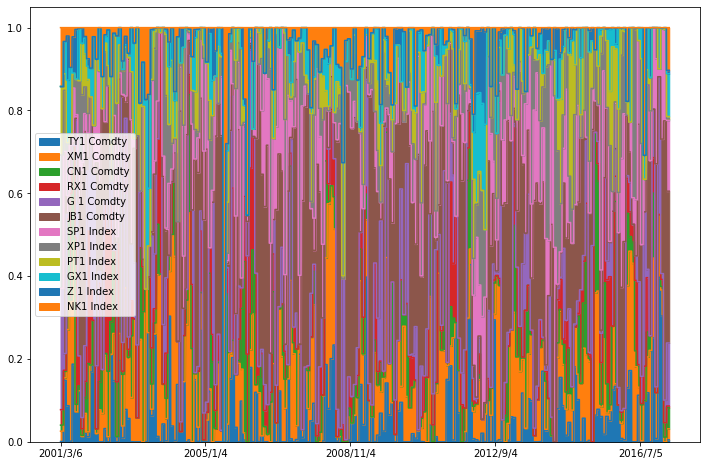}
  \caption{The allocation of the assets in the HPCA portfolio.}
  \label{Wgt_HPCA}
\end{figure}

\begin{figure}
  \centering
  \includegraphics[width=0.9\textwidth]{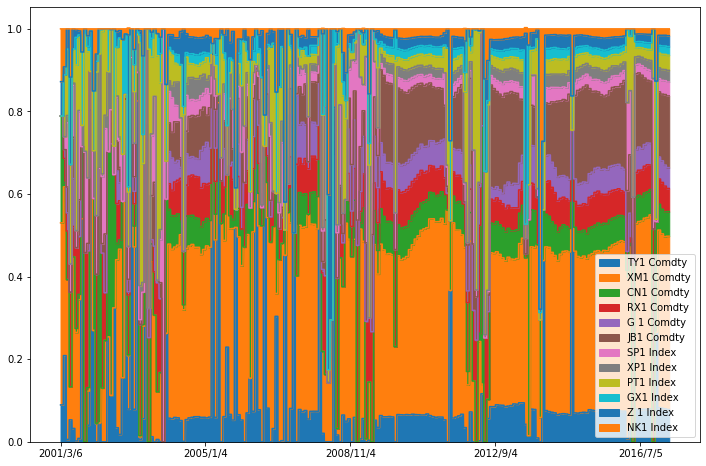}
  \caption{The allocation of the assets in the SPCA portfolio.}
  \label{Wgt_SPCA}
\end{figure}

Table~\ref{Avgwgt} shows the time averaged weight coefficients of the portfolios in terms of assets and type of assets, respectively.
The corresponding sequences of the weights of the PCA, HPCA and SPCA portfolios are shown in Figures \ref{Wgt_PCA}-\ref{Wgt_SPCA}.
We can see that the weights of equity futures are higher in the order of PCA, SPCA, and HPCA.

\begin{table}[]
\caption{Averaged weights of the portfolios.}
\label{Avgwgt}
\centering
\begin{tabular}{|l|r|r|r|}
\hline
\multicolumn{1}{|c|}{} & \multicolumn{1}{c|}{PCA~[\%]} & \multicolumn{1}{c|}{HPCA~[\%]} & \multicolumn{1}{c|}{SPCA~[\%]} \\ \hline
TY1                    & 17.04                    & 6.40                      & 11.63                     \\ \hline
XM1                    & 23.19                    & 10.88                     & 31.22                     \\ \hline
CN1                    & 14.49                    & 7.37                      & 6.96                      \\ \hline
RX1                    & 9.14                     & 6.17                      & 8.76                      \\ \hline
G 1                    & 8.69                     & 7.28                      & 3.87                      \\ \hline
JB1                    & 6.57                     & 25.32                     & 9.52                      \\ \hline
SP1                    & 5.57                     & 10.06                     & 5.75                      \\ \hline
XP1                    & 4.17                     & 9.32                      & 2.32                      \\ \hline
PT1                    & 4.00                     & 5.78                      & 9.37                      \\ \hline
GX1                    & 3.23                     & 4.87                      & 3.26                      \\ \hline
Z 1                    & 1.98                     & 2.09                      & 3.82                      \\ \hline
NK1                    & 1.92                     & 4.45                      & 3.52                      \\ \hline \hline
Bond                   & 79.12                    & 63.42                     & 71.96                     \\ \hline
Equity                 & 20.88                    & 36.58                     & 28.04                     \\ \hline
\end{tabular}
\end{table}
\section{Conclusion}
%
%
We proposed the Schr\"{o}dinger risk diversification portfolio which applies Schr\"{o}dinger PCA to extract risk factors.
We perform experiments on multi-asset market data to evaluate the proposed portfolio. Compared with various portfolios, the proposed portfolio demonstrates a higher risk-adjusted returns.

In the empirical analysis, we used the harmonic oscillator potential as the potential function and assumed the spatial arrangement among the assets are evenly spaced.
In general, when we conduct Schr\"{o}dinger PCA, there is room for selecting the potential function and making the spatial arrangement unreasonable.
Also, we can extend our proposed method to complexifying the time series of returns of each asset by Hilbert transform as in \cite{uchiyama2019complex}.
The future task is to verify the effectiveness of the proposed method developed in these directions.

\bibliographystyle{unsrt}  

\end{document}